# Growth and Thermo-driven Crystalline Phase Transition of Metastable Monolayer 1T'-WSe$_2$ Thin Film

Wang Chen[1], Xuedong Xie[1], Junyu Zong[1], Tong Chen[1], Dongjin Lin[1], Fan Yu[1], Shaoen Jin[1], Lingjie Zhou[1], Jingyi Zou[1], Jian Sun[1,2], Xiaoxiang Xi[1,2], Yi Zhang[1,2]*

[1]National Laboratory of Solid State Microstructure, School of Physics, Nanjing University, Nanjing, 210093, China

[2]Collaborative Innovation Center of Advanced Microstructures, Nanjing University, Nanjing, 210093, China

*Email: zhangyi@nju.edu.cn

Two-dimensional (2D) transition metal dichalcogenides MX$_2$ (M = Mo, W, X = S, Se, Te) attracts enormous research interests in recent years. Its 2H phase possesses an indirect to direct bandgap transition in 2D limit, and thus shows great application potentials in optoelectronic devices. The 1T' crystalline phase transition can drive the monolayer MX$_2$ to be a 2D topological insulator. Here we realized the molecular beam epitaxial (MBE) growth of both the 1T' and 2H phase monolayer WSe$_2$ on bilayer graphene (BLG) substrate. The crystalline structures of these two phases were characterized using scanning tunneling microscopy. The monolayer 1T'-WSe$_2$ was found to be metastable, and can transform into 2H phase under post-annealing procedure. The phase transition temperature of 1T'-WSe$_2$ grown on BLG is lower than that of 1T' phase grown on 2H-WSe$_2$ layers. This thermo-driven crystalline phase transition makes the monolayer WSe$_2$ to be an ideal platform for the controlling of topological phase transitions in 2D materials family.

Quantum spin Hall (QSH) effect is a topologically protected state with helical edge channels for one-way spin transport without dissipation, which can be realized in a two-dimensional (2D) topological insulator[1,2]. The QSH effect was first observed in HgTe quantum well system[3]. After that, amounts of 2D materials was proposed to host such QSH state, such as gapped graphene[4], silicene, germanene[5], stanene[6], etc. Recently, monolayer transition metal dichalcogenides (TMDCs) MX$_2$ (M = Mo, W, X = S, Se, Te) with monoclinic 1T' structure was suggested to be a 2D topological insulator[7]. Later, the topological band structures of monolayer 1T'-WTe$_2$ and 1T'-WSe$_2$ were characterized by using angle-resolved photoemission spectroscopy and scanning-tunneling spectroscopy[8,9,10]. By constructing the monolayer 1T'-WTe$_2$ with h-BN substrate and capping layer as a sandwich heterostructure, the QSHE that can survive under 100 K was also observed recently[11].

As a member of 2D materials family, TMDCs MX$_2$ has distinct electronic structures with various crystalline structures, thus attracts enormous research interests in recent years. The MX$_2$ with hexagonal 2H crystalline structure is a semiconductor with indirect bandgap. When the thickness of 2H-MX$_2$ is reduced to monolayer limit, such indirect bandgap will transit into direct bandgap[12,13,14], and a giant spin splitting up to ~0.5 eV is formed in the top of valence band[15,16]. With the 2H crystalline structure transforming into orthorhombic Td structure, MoTe$_2$ and WTe$_2$ were confirmed to be a type-II Weyl semimetal[17,18,19,20]. As a QSH insulator, only the WTe$_2$ was predicted to be most stable in 1T' phase[7,21]. For the MoTe$_2$, the Td to 1T' phase transition can be driven by temperature or dimension[22], and the 2H to 1T' phase transition can be driven by electrostatic doping[23]. So far, the structure phase transition of monolayer WSe$_2$ has not been reported since the total energy of 1T'-WSe$_2$ is predicted much higher than that of 2H-WSe$_2$ phase. Here we successfully obtained the monolayer 1T'-WSe$_2$ film on bilayer graphene (BLG) substrate using molecular beam epitaxial (MBE) method. Combining the *in-situ* scanning tunneling microscopic (STM) studies, we found that the 1T' phase of WSe$_2$ is metastable and can transform into stable 2H phase under high-temperature annealing. Since the 1T'-WSe$_2$ was predicted to host larger topological bandgap than 1T'-WTe$_2$[7], such structure phase transition of WSe$_2$ enables it to be an ideal platform to study the topological phase transition and devices based on the TMDCs.

Figure 1a-1f show the comparison of the crystalline structures between 2H-WSe$_2$ and 1T'-WSe$_2$ monolayers. From the top-view (Figure 1b), the monolayer 2H-WSe$_2$ show a honeycomb structure similar to graphene. From the side-view (Figure 1c), the Se atoms of top layer are sited vertically on the those of bottom layer, consisting a 2H-WSe$_2$ monolayer with a layer of W atoms insert in the middle as a sandwich structure. Comparatively, the in-plane orientation of top Se layer is rotated by 180° relative to the bottom Se layer in 1T'-WSe$_2$ (Figure 1d &1e), meanwhile the adjacent two W arrays become closer to each other, which makes one array of top Se atoms become higher than the adjacent one (Figure 1f). Such deformation

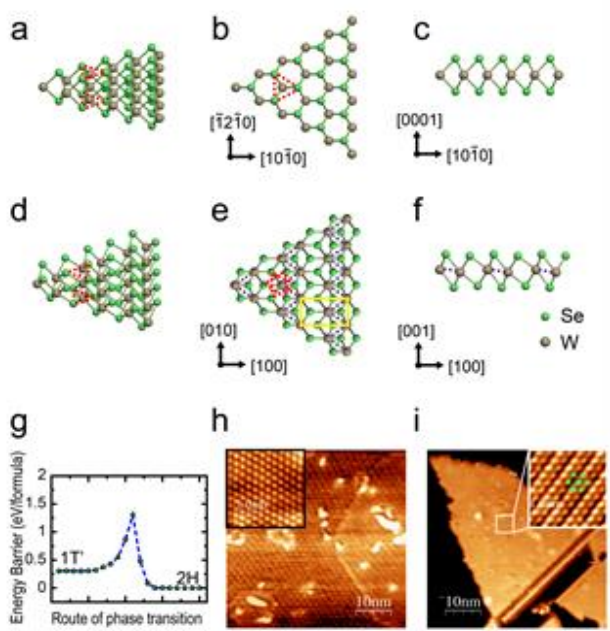

**Figure 1 (a)** 3D view, **(b)** top view and **(c)** side view of 2H-WSe$_2$ monolayer. **(d)** 3D view, **(e)** top view and **(f)** side view of 1T'-WSe$_2$ crystalline. **(g)** Energy Barrier for the 1T' to 2H structure phase transition of monolayer WSe$_2$. **(h)&(i)** STM image of BLG substrate and grown WSe$_2$ films at substrate temperature of 250°C. Insets are the atomic resolution images. The green dots indicate the position of Se atoms of top layer. Scanning parameters for STM: $V_{tip}$ = 1 V, I = 100 pA.

makes monolayer 1T'-WSe$_2$ host a rectangle unit cell rather than the hexagonal one in monolayer 2H-WSe$_2$. Figure 1g shows the difference of total energy between 2H and 1T' phase of WSe$_2$ monolayer. The 2H phase is most stable with lowest total energy, while the 1T' phase is metastable with local minimum energy.

In previous report, monolayer MX$_2$ can be grown on bilayer graphene (BLG) substrate using MBE method[9, 10, 13, 14, 24, 25, 26, 27]. Here we also prepared the BLG by annealing 6H-SiC(0001) to 1300°C for 40~80 cycles[28]. Figure 1h shows the STM image of a BLG substrate and its atomic resolution, indicating that such substrate is ready for the film growth. We co-deposited W and Se with flux ratio ~ 1:10. In Se rich growth condition, the growth rate of the WSe$_2$ film is only dominated by the W flux, which was about 0.15 monolayer per minutes in our case (Supplementary Information D). Even though the total energy of 1T'-WSe$_2$ is higher than that of 2H phase (Figure 1g), a 1T'-WSe$_2$ monolayer can be formed on the substrate when we controlled the substrate temperature at 250°C during the growth. Figure 1i show a STM image of a 1T'-WSe$_2$ domain with significant stripe-like topography. The insert atomic-resolution image indicates that these stripes in owing to the height differences of the top Se arrays as shown in Figure 1f, from which we can get the lattice constant of 1T'-WSe$_2$ is about a = 5.8Å, b = 3.3Å. With statistics of large-scale STM images of such grown films, the ratio of 1T'-WSe$_2$ domains to 2H ones is about 1:1 (see Supplementary Information A).

Figure 2a-2c show images of the monolayer WSe$_2$ films grown with different substrate temperatures. When we increased the substrate temperature to 300°C during the growth, the average size of WSe$_2$ domains became larger than that grown at 250°C. Figure 2a shows the coexistence of the 1T' domains and 2H domains in an STM image. The 1T' domain has a significant stripes feature, while the 2H domain has a flat surface. An atomic resolution image of 2H domain in Figure 3c shows that the heights of the Se atoms in the top layer is exactly the same, indicating the distinct crystalline structure of 2H phase to the 1T' phase. Besides that, we note that the edges of 2H domains are always saturated with adatoms, while the edges of 1T' domains are very clean and sharp. The difference of these

2 / 7

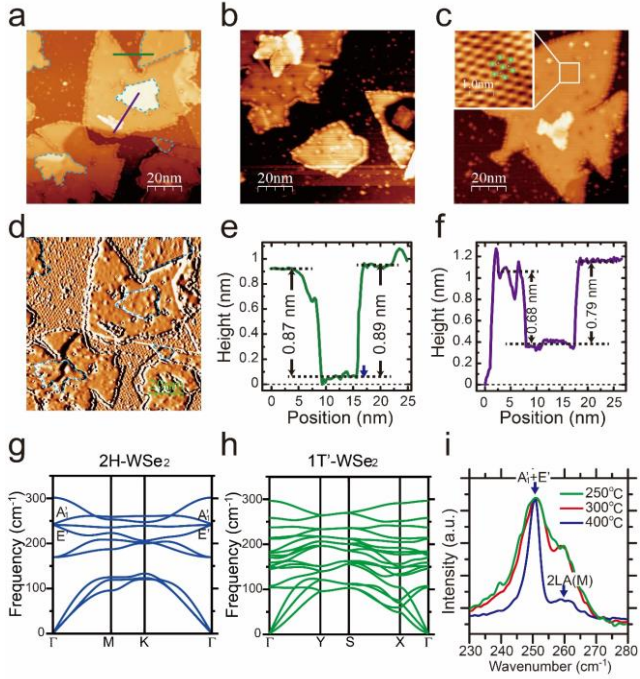

**Figure 2.** **(a)-(c)** Images of grown WSe$_2$ at substrate temperature of 300°C (a), 350°C (b), and 400°C (c), respectively. **(d)** Derivative images of (a). The blue dashed contours depict the domains of 1T' phase. **(e) & (f)** Height profiles along the green and purple lines in (a), respectively. **(g) & (h)** Calculated phonon spectra of 2H and 1T' WSe$_2$ monolayer, respectively. **(i)** Raman spectra of the films grown at 250°C, 300°C and 400°C, respectively. Scanning parameters for STM: $V_{tip}$ = 1 V, I = 100 pA.

two types of domains can be clearly identified in derivative image in Figure 2d. The domains of 1T' phase with stripes feature are depicted by the blue dashed contours in Figure 2a and 2d. Figure 2b shows that when we further increased the substrate temperature to 350°C during the growth, the areas of 1T' phase dramatically reduced and the only the areas of 2H phase can be observed. With higher substrate temperature of 400°C, unitary 2H-WSe$_2$ film with large triangle domains were grown on BLG in Figure 2c.

Figure 2e and 2f show the height profiles along the green and purple line in Figure 2a. In Figure 2e, the height of 1T' phase (~ 0.89 nm) is slightly higher than that of 2H phase (~ 0.87 nm) for the domains grown on the BLG substrate. However, for the domains grown on the 2H WSe$_2$ monolayer, the height of 1T' phase (~ 0.79 nm) is obviously higher than that of 2H phase (~ 0.68 nm). The higher height profile of the film on BLG than island on

WSe$_2$ monolayer implies that the Van de Waals bonding in the interface between BLG and WSe$_2$ is weaker than the interface between WSe$_2$ layers. The height of 2H-WSe$_2$ island on the monolayer domain is exactly coincide with the lattice constant c of 2H-WSe$_2$. One reason for the higher height profile of 1T' phase in our STM image may be the conductive nature of 1T' phase. The other reason may be that in 1T' phase, the adjacent two W arrays become closer to each other, which makes one array of top Se atoms becomes higher than the adjacent array (Figure 1f), such deformation makes monolayer lattice of 1T' phase become higher than that of 2H phase in STM images.

The 1T' phase also shows distinct spectrum from the 2H phase in Raman measurements due to its different structure. Figure 2i shows the Raman spectra of monolayer WSe$_2$ films grown with different substrate temperatures of 250 °C (green line), 300 °C (red line) and 400 °C (blue line), respectively. The calculated phonon spectra of 2H and 1T' phases are shown in figure 2g and 2h, respectively. For the 2H phase (blue line in Figure 2i), the Raman spectrum is coincided with previous report, with a mixture of A$_1$' and E' mode at ~250 cm$^{-1}$, and a weaker second-order Raman mode [2LA(M)] at ~260 cm$^{-1}$ [29, 30, 31]. But for the 1T' phase, the peak of A$_1$' and E' mode becomes much broaden for the green and red lines in figure 2i, this can be owing to that the A$_1$' and E' mode splits into three branches at the Γ point in Figure 2g.

Since the total energy of 1T' phase is higher than that of 2H phase for monolayer WSe$_2$, when we performed the higher temperature post-annealing procedures on 1T' phase, the thermo-driven 1T' to 2H phase transition was observed. Here we first prepared a 1T'-2H mixed WSe$_2$ films with a coverage of ~ 1.3 monolayer, which has monolayer 1T'-WSe$_2$ domains on both the BLG and 2H-WSe$_2$ layers simultaneously (Figure 3g). Then we performed the post-annealing with various temperature from 250°C to 800°C, step by step, with 25°C increasing for each step. Each annealing procedure was carried out for 20 minutes, which is longer enough for the phase transition (Supplementary C). During the annealing procedures, the Se flux was kept opening to avoid possible desorption of Se atoms and the formation of Se vacancy defects. Figure 3a shows the RHEED pattern in BLG substrate with the incident electron beam along the $[10\bar{1}0]$ direction of graphene lattices. Figure 3b shows the RHEED pattern after the film was annealed at 275°C. The yellow arrow indicates the (1×1) diffraction from the top Se layer of 1T'-WSe$_2$/2H-WSe$_2$, both of them has an in-plane hexagonal lattice in the top-view. The space between the (1×1) spots of WSe$_2$ film is about 3/4 to the space between BLG (1×1) diffraction spots in Figure 3a, indicating the lattice constant of 2H WSe$_2$ is ~ 4/3 times to that of BLG, coinciding with our STM results. The characterized diffraction spots of 1T' phase can be clearly observed, indicating that the 1T' phase is still there. Combining with their weak intensity, we believe these diffraction spots are owing to the height difference of top Se atoms in 1T'-WSe$_2$, which induce a double crystalline period comparing to the 2H phase along the [100] direction (yellow rectangle in Figure 1e). These distinct diffraction spots can be used to identify the 1T' structure in WSe$_2$ films[9, 10]. Figure 3c shows that after annealing at 600°C, the diffraction spots from 1T' phase is totally disappeared, only the 2H phase is exist, implying all the 1T' phase transformed into 2H phase during the annealing procedure. Figure 3d shows that when the annealing temperature is over 750°C, the RHEED pattern from 2H phase is completely disappeared. The matrix-like diffraction spots indicate that the film was decomposed, and the residual W atom concentrated together as three-dimensional clusters.

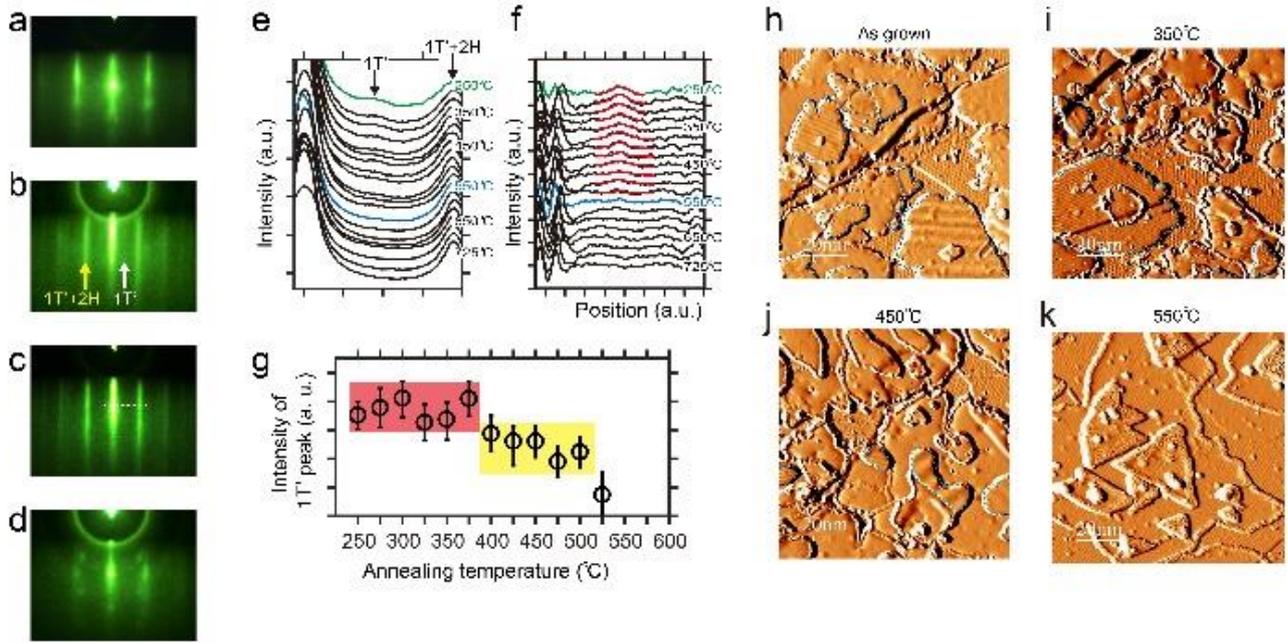

**Figure 3**. **(a) & (b)** RHEED pattern of BLG substrate and grown WSe$_2$ film at substrate temperature of 250°C, respectively. **(c) & (d)** RHEED pattern of the film after annealing at 600°C and 750°C, respectively. **(e)** Intensity distribution profiles along the white dashed line demonstrated in RHEED image (c) for the films after annealing at temperature ranging from 275°C to 725°C, with 25°C steps. **(f)** Intensity distribution profiles after deducting the background from (e). The red lines are the Gaussian fitting lines of the diffraction peaks of the 1T' phase. **(g)** Intensities of the fitting results of the 1T' diffraction peaks in (f). **(h)-(k)** Derivative STM images of the WSe$_2$ film grown at 250°C, and post annealing at 350°C, 450°C, 550°C for 20 minutes, respectively. The 1T' domains are indicated by the blue dashed contours. Scanning parameters for STM: $V_{tip}$ = 1 V, I = 100 pA.

To better determine the phase transition temperature accurately, we extract the intensity distribution lines of diffraction spots along the white dashed line demonstrated in Figure 3c for each RHEED pattern (Figure 3e). The intensity of the diffraction peak from 1T' phase is rather weak comparing to the other peaks. For quantitatively characterizing the intensities of 1T' diffraction peaks, the (1×1) diffraction peaks were normalized in Figure 3e. To distinguish the evolution of these weak peaks of 1T' phase, we deduct other peaks as background by using Gaussian fitting in Figure 3f. The red lines are the Gaussian fitting lines of the 1T' diffraction peaks. We farther extracted the intensities of 1T' diffraction peaks from the fitting results and plot them on Figure 3g. Here we note when the annealing temperature is bellow 375°C, the intensity of 1T' peak does not fade (indicated by the red region in Figure 3g), implying that these annealing temperatures could not induce the 1T' to 2H phase transition. Figure 3i shows a STM image taken after annealing at 350 °C. We can see that the 1T' domains still survive. When the annealing temperature is over 400 °C, there is a distinct dropping of peak intensity indicated by the yellow region in Figure 3g. Figure 3j shows a STM image taken after annealing at 450°C. We found that all the 1T' domains grown on BLG transform into 2H phase, but the 1T' domains on the 2H-WSe$_2$ layers still survive. Therefore, the structure transition of 1T' domains grown on BLG reduces the intensities of 1T' peaks distinctly in the yellow region of Figure 3g. When the annealing temperature is over 500 °C, the diffraction peak of 1T' phase was totally disappeared, implying that all the 1T' phase transformed into 2H structure. This phase transition cannot be reversed when we cool down the sample back to the room-temperature, which means the 2H is a stable phase while the 1T' phase is metastable.

Combining analysis of the RHEED patterns and STM images in the post-annealing experiments, we found that the 1T' to 2H transition temperature of 1T'-WSe$_2$ domains directly grown on BLG is ~400 °C (Supplementary Information B), which is much lower than that of 1T'-WSe$_2$ domains on 2H layers (~525 °C). This difference implying that the stability of 1T' phase is also dependent on the interface between substrate. From the analysis of the domains heights that were discussed above, the interaction between WSe$_2$ and BLG substrate is weaker than that between WSe$_2$ layers, which may be the reason that 1T' phase grown on BLG is less stable than that on 2H layers.

The phase transition temperature from 1T' to 2H for monolayer WSe$_2$ film on BLG is apparently higher than the growth temperature of a unitary 2H-WSe$_2$ monolayer, which is at least 350°C. This difference can be owing to the non-equilibrium growth mode in the MBE, thus the temperature for forming a stable phase is mainly determined by the energy difference between metastable 1T' phase and stable 2H phase. For the phase transition from a metastable 1T' phase to a stable 2H phase, the thermal agitation energy must be higher enough for crossover the energy barrier between these two phases, which require a higher annealing temperature than that for growth of stable 2H phase.

In summary, we successfully grew metastable 1T'-monolayer WSe$_2$ films on BLG substrate using MBE method. At higher growth temperature, we can get pure 2H-WSe$_2$ monolayer WSe$_2$. The 1T' and 2H-WSe$_2$ monolayer show distinctly different crystalline structures in STM images, and also the different characterizations in RHEED and Raman measurements. The growth of 1T'/2H phase-mixture film could be used for fabricate the lateral 2D metal-semiconductor heterojunction. And the thermo-driven phase transition from 1T' to 2H in monolayer WSe$_2$ open a new route for the control of structure and electronic properties in 2D materials.

**Methods**

The growth of the WSe$_2$ films was performed in a combined MBE-STM ultra-high vacuum (UHV) system with base pressure of ~ 1.5×10$^{-10}$ mbar. The W flux was produced from a high purity (99.95%) tungsten rod using an electron-beam heating evaporator with flux monitor function. The high purity Se (99.9995%) was evaporated from a standard Knudsen cell. Both the flux of W and Se was calibrated by depositing them on a Si(111)-7×7 reconstructive surface using *in-situ* reflection high-energy diffraction (RHEED) and STM monitoring. The temperature of the sample was measured with a Photrix pyrometer with temperature ranging from 135 °C to 2400 °C. The STM is a Pan-style one and performed at room temperature. Raman scattering measurements were performed using a home-built confocal microscope equipped with a grating spectrometer and a liquid-nitrogen-cooled charge coupled device from Princeton Instruments. Unpolarized spectra were taken with 532 nm laser excitation in the back-scattering geometry at ambient conditions, with the incident laser power limited to below 1 mW. The calculations were performed with Vienna *ab-initio* Simulation Package (VASP)[32]. The phonon spectrum and the active Raman modes of 1T' and 2H phase were calculated by using Phonopy package with a 2x2x1 supercell[33]. The transition barrier between the two phases was calculated using Variable-Cell Nudged-Elastic-Band (VCNEB) method implemented in USPEX code[34] combining with VASP. For preventing the contamination and oxidization of the film during the transferring, an ~20 nm Se capping layer was deposited on the surface before taking the sample out of the UHV chamber.[35]

**Acknowledgements**

This work was supported by the National Key R&D Program of China (No.2018YFA0306800), National Natural Science Foundation of China (Grant Nos. 11714154, 11790311 and 11774151), the Fundamental Research Funds for the Central Universities (No.020414380110) and the Program for High-Level Entrepreneurial and Innovative Talents Introduction, Jiangsu Province.

**Author contributions**

Y. Z. conceived the experiments, and wrote the paper with suggestions and comments by J. S. and X. Xi. W. C. led the MBE growth and STM characterization of the samples with the assistance of X. Xie, J. Zong, F. Y., S. J., L. Z., and J. Zou. T. C. and J. S. performed the calculations of phonon spectra and energy barrier for phase transition. D. L. and X. Xi performed the Raman spectrum measurements.

**Additional Information**
The authors declare no competing interests.

# Supplementary Informaion for

# Growth and Thermo-driven Crystalline Phase Transition of Metastable Monolayer 1T'-WSe$_2$ Thin Film

Wang Chen, Xuedong Xie, Junyu Zong, Tong Chen, Dongjin Lin, Fan Yu, Shaoen Jin, Lingjie Zhou, Jingyi Zou, Jian Sun, Xiaoxiang Xi, Yi Zhang*

*Email: zhangyi@nju.edu.cn

### A: Ratio of 1T' to 2H phase of monolayer WSe$_2$ grown on BLG

We grew some WSe$_2$ sample on BLG with coverage ~0.4 ML and substrate temperature about 250°C. Then we obtained numerous of STM images. The 1T' and 2H domains can be easily distinguished by using negative tip bias of -1.5V, which is in the energy gap of 2H-WSe$_2$ and make it be almost transparent in STM images. We measure the area of the 1T' domains and 2H domains of each STM images as shown in Figure S1. Then we got the ratio of 1T' to 2H phase is about 1:1.

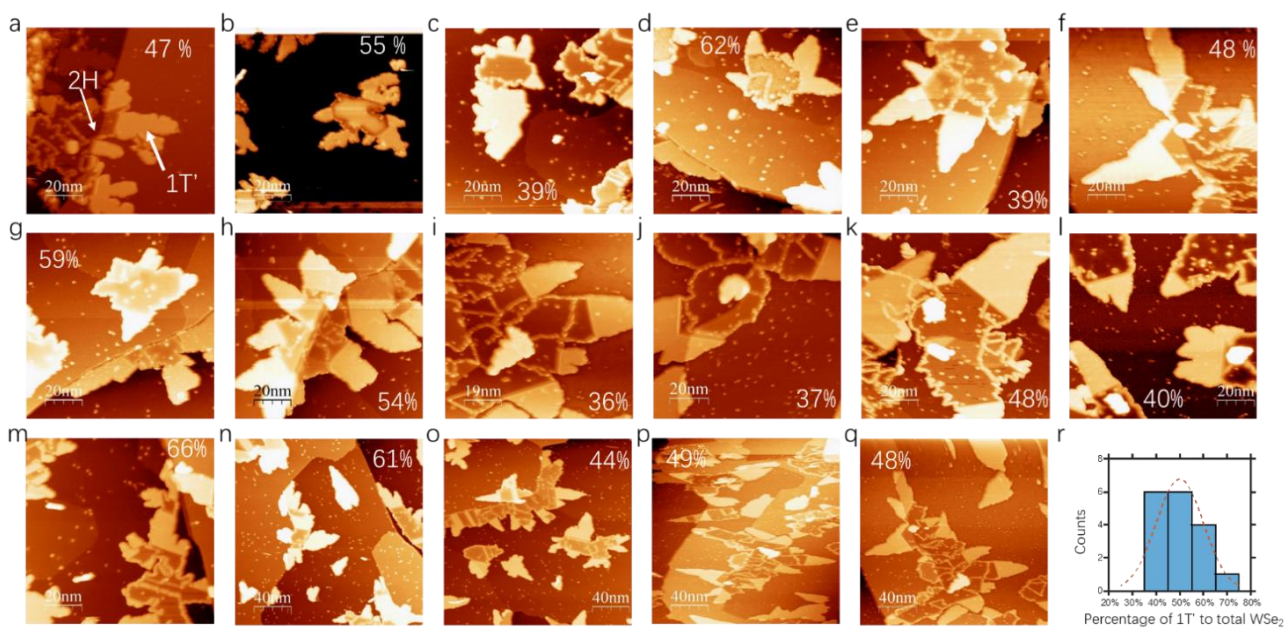

**Figure S1. (a)-(q)** STM images of WSe$_2$ film grown on BLG at substrate temperature of 250 °C. The inset percentiles are the percentage of 1T' phase to the total WSe$_2$ area in each image. **(r)** Histogram of the percentiles in each image.

## B: 1T' to 2H phase transition of monolayer WSe$_2$ grown on BLG

We prepared a WSe$_2$ on BLG with substrate temperature of ~250°C and coverage of ~0.3 ML. This low coverage ensure that the monolayer 1T'-WSe$_2$ only formed on the BLG substrate, no 1T'-WSe$_2$ islands formed on the top of monolayer WSe$_2$ layer. Then we did the 400°C post-annealing. Both the STM images and RHEED patterns show that all the 1T' domains transit into 2H phase (Figure S2). Thus we conclude that for the 1T'-WSe$_2$ monolayer grown on BLG, the 1T' to 2H phase transition temperature is below 400°C, which is lower than the phase transition temperature of 1T'-WSe$_2$ grown on 2H phase that was discussed in main text.

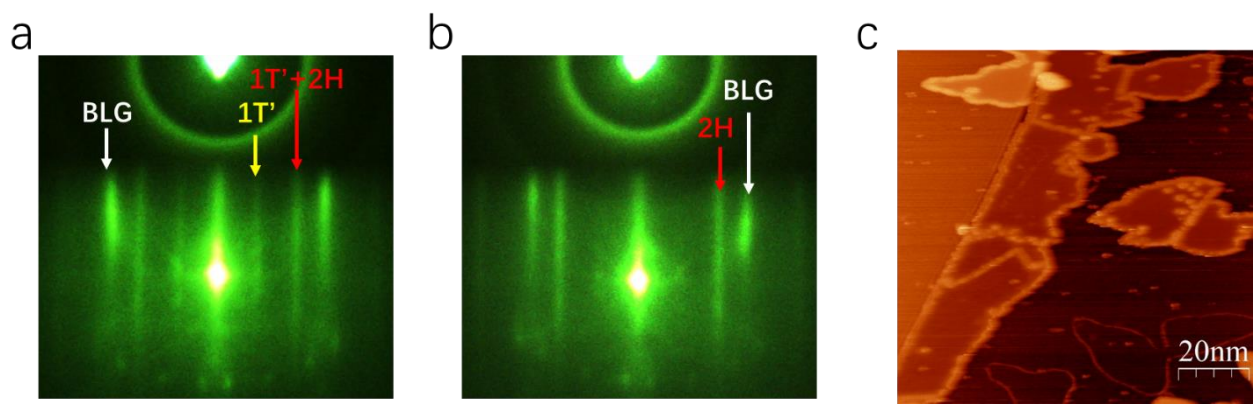

**Figure S2. (a)** RHEED pattern of a sample with monolayer 1T'-WSe$_2$ only on the BLG substrate. **(b)** After post-annealing the sample at 400°C for 20 min, the diffraction spots of 1T' phase was totally disappeared. **(b)** STM image show that all the 1T' phase transited into 2H phase after the post-annealing

## C: The duration time for post-annealing procedures

We prepared a sample with ~0.3 ML coverage of 1T'-WSe$_2$, which have monolayer 1T' domains only grown on BLG but not on 2H-WSe$_2$ layers (Figure S3a). We first did the post-annealing at 350 °C for 1 hour, this temperature is below the phase transition temperature. We monitored the RHEED image and found that the intensity of diffraction spots from 1T' phase did not decay with the time for post-annealing (Figure S3b-d). Once we increase the annealing temperature up to 400°C, the 1T' to 2H phase transition immediately happens with in few seconds (Figure S3 e&f). Thus we conclude that the 20 minutes post-annealing time mentioned in the main text is longer enough for the totally 1T' to 2H phase transition of WSe$_2$ monolayer.

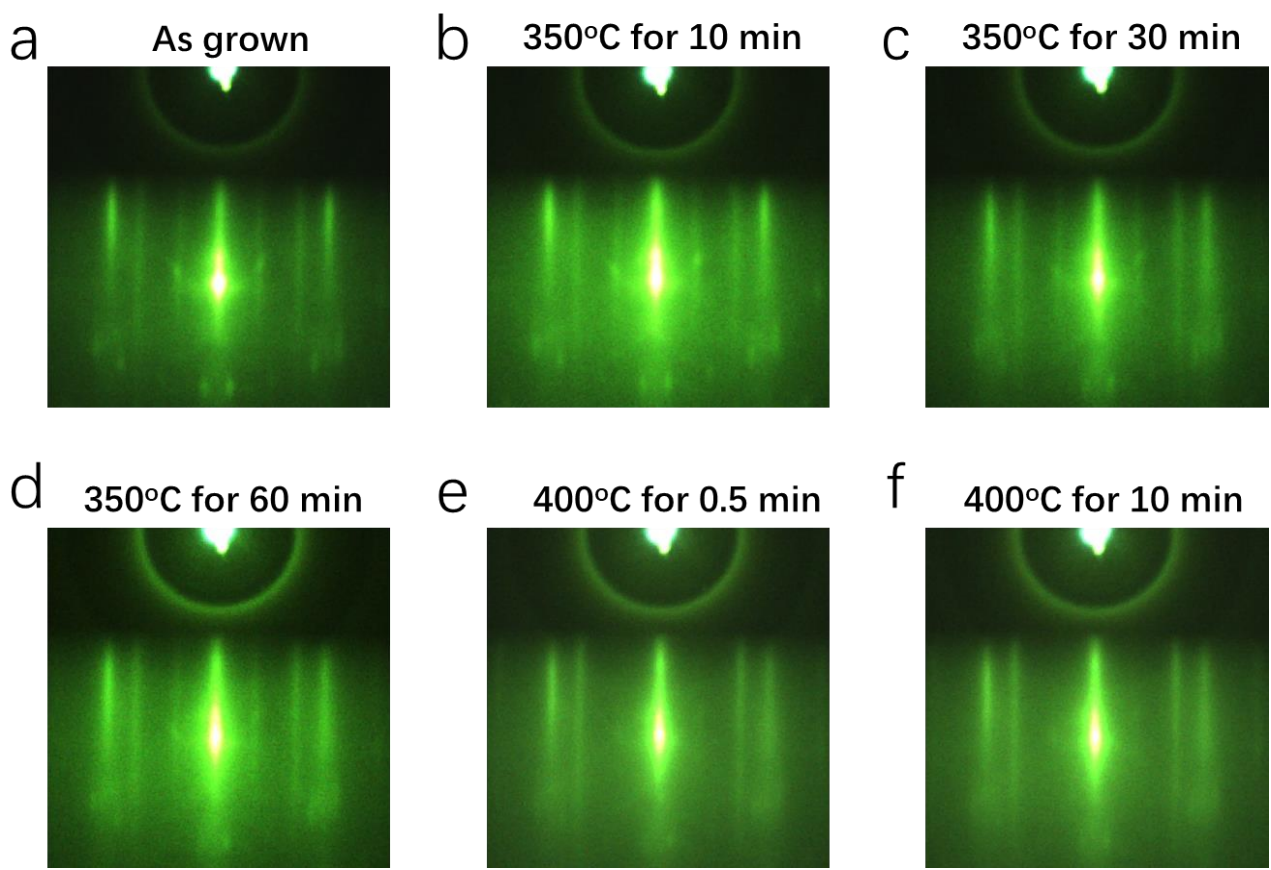

**Figure S3.** RHEED patterns of a sample with monolayer 1T'-WSe$_2$ only on the BLG substrate: **(a)** RHEED pattern just taken after the growth. **(b)-(f)** RHEED pattern after different post-annealing procedures.

*D: Different Se flux used for the growth of thin films*

We did the WSe$_2$ growth with different Se fluxes. The temperatures of Se source are ranging from 130°C to 160°C. The corresponding fluxes are from ~1 ML/min to ~10 ML/min. Since the substrate temperature is ~250°C, which is significantly higher than the temperature of Se source and molecular beam, the Se atoms cannot be solely adsorbed on the chemically inert surface of BLG or WSe$_2$. Notably, the substrate temperature is lower than the temperature of W flux ( > 2000°C). Therefore, the W atoms can be adsorbed on the surface of relatively cold substrate. Since the Se flux is excessive, once a W atom is adsorbed on the surface, the Se molecules will reactive with the adsorbed W atom and forms WSe$_2$ film. Thus $T_W > T_{sub} > T_{Se}$ growth dynamics was also reported in the MBE growth of topological insulator Bi$_2$Se$_3$ and Bi$_2$Te$_3$[1]. The growth rate of the WSe$_2$ film is only dominated by the flux W in such growth conditions. In Figure S4, we found the film morphology has no obviously difference.

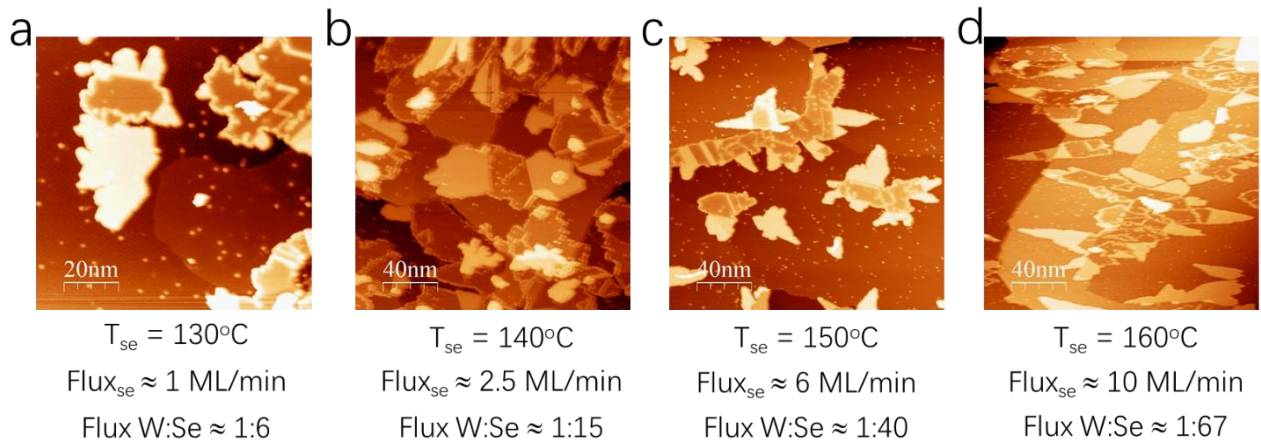

| | | | |
|---|---|---|---|
| $T_{se}$ = 130°C | $T_{se}$ = 140°C | $T_{se}$ = 150°C | $T_{se}$ = 160°C |
| Flux$_{se}$ ≈ 1 ML/min | Flux$_{se}$ ≈ 2.5 ML/min | Flux$_{se}$ ≈ 6 ML/min | Flux$_{se}$ ≈ 10 ML/min |
| Flux W:Se ≈ 1:6 | Flux W:Se ≈ 1:15 | Flux W:Se ≈ 1:40 | Flux W:Se ≈ 1:67 |

**Figure S4.** STM images of the WSe$_2$ films grown with different flux of Se.